\documentclass[aps,prl,groupedaddress,twocolumn,showpacs,10pt]{revtex4-1}
\usepackage{graphicx,color}
\usepackage{bm}
\usepackage{amssymb,amsmath,amsfonts, mathrsfs}

\newcommand{\be}{\begin{equation}}
\newcommand{\ee}{\end{equation}}
\newcommand{\ba}{\begin{array}}
\newcommand{\ea}{\end{array}}
\newcommand{\bqa}{\begin{eqnarray}}
\newcommand{\eqa}{\end{eqnarray}}

\begin{document}


\title{Understanding $X(3862)$,  $X(3872)$, and $X(3930)$  in a Friedrichs-model-like scheme}


\author{Zhi-Yong Zhou}
\email[]{zhouzhy@seu.edu.cn}
\affiliation{School of Physics, Southeast University, Nanjing 211189,
P.~R.~China}
\author{Zhiguang Xiao}
\email[]{xiaozg@ustc.edu.cn}
\affiliation{Interdisciplinary Center for Theoretical Study, University of Science
and Technology of China, Hefei, Anhui 230026, China}


\date{\today}

\begin{abstract}
We developed a Friedrichs-model-like scheme in studying the hadron
resonance phenomenology and present that the hadron resonances might
be regarded as the Gamow states produced by a Hamiltonian  in which
the bare discrete state is described by the result of usual quark
potential model and the interaction part is described
by the quark pair creation model. In a  one-parameter
calculation, the $X(3862)$,  $X(3872)$, and $X(3930)$ state could be
simultaneously produced with a quite good accuracy by coupling the
three P-wave states, $\chi_{c2}(2P)$, $\chi_{c1}(2P)$, $\chi_{c0}(2P)$
predicted in the Godfrey-Isgur model to  the $D\bar D$, $D\bar D^{*}$, $D^*\bar D^*$
continuum states. At the same time, we predict that the $h_c(2P)$
state is at about 3902 MeV with a pole width of about 54 MeV. In this
calculation, the $X(3872)$ state has a large compositeness.  This scheme
may shed more light on the long-standing problem about the general
discrepancy between the prediction of the quark model and the observed
values, and it may also provide reference for future search for the
hadron resonance state.
\end{abstract}


\maketitle
As regards the charmonium spectrum above the open-flavor thresholds, general
discrepancies between the predicted masses in the quark potential
model and the observed values have been highlighted for several years.
Typically, among the P-wave $n^{2s+1}L_J=$  $2^3P_2$, $2^3P_1$,
$2^3P_0$, and $2^1P_1$ states, the $X(3930)$, discovered by the Belle
Collaboration\cite{Uehara:2005qd}, is now assigned to $\chi_{c2}(2P)$
charmonium state though its mass is about 50 MeV lower than the
prediction in the quark potential
model~\cite{Eichten:1978tg,Godfrey:1985xj,Barnes:2005pb}. The
properties of the other  P-wave states have
not been firmly determined yet. The $X(3872)$ was first observed in
the $B^\pm\rightarrow K^\pm J/\psi\pi^+\pi^-$ by the Belle
Collaboration in 2003~\cite{Choi:2003ue}. Although its quantum number
is $1^{++}$,  the same as the $\chi_{c1}(2P)$,  the pure charmonium
interpretation was soon given up for the difficulties in explaining its
decays. The pure molecular state explanation of $X(3872)$ also
encounters difficulties in understanding its radiative decays. So its
nature remains to be obscure up to now. As for the $\chi_{c0}(2P)$
state,  the $X(3915)$ is assigned to it several years ago, but this
assignment is questioned for the mass splitting between
$\chi_{c2}(2P)$ and $\chi_{c0}(2P)$, and its dominant decay
mode~\cite{Guo:2012tv,Olsen:2014maa}. In Ref.\cite{Zhou:2015uva},
analyses of the angular distribution of $X(3915)$ to the final
leptonic and pionic states also support the possibility of being a $2^{++}$ state,
which means that it might be the same tensor state as the $X(3930)$.
Very recently, the Belle Collaboration announced a new result about
the signal of $X(3862)$ which could be a candidate for the
$\chi_{c0}(2P)$~\cite{Chilikin:2017evr}.  The $2^1P_1$ state has not
been discovered yet.  These puzzles have been discussed exhaustively
in the literatures~(see refs.~\cite{Chen:2016qju,Esposito:2016noz,Lebed:2016hpi} for example), but a consistent description is still missing.

In this paper, we adopt the idea of Gamow states and the solvable
extended Friedrichs model developed
recently~\cite{Xiao:2016dsx,Xiao:2016wbs,Xiao:2016mon}, usually
discussed in the pure mathematical physics literature, to study the
resonance phenomena in the hadron physics, in particular the
charmonium spectrum. Using the eigenvalues and
wave functions for mesons in the Godfrey-Isgur~(GI) model~\cite{Godfrey:1985xj}
as input and modelling the interaction by the quark pair production~(QPC)
model, we found that the  first excited $2^{++}$, $1^{++}$,  and $0^{++}$ charmonium states could
be reproduced with good accuracy in an one-parameter
calculation, and the  mass and width of $1^{+-}$ state are also obtained as a
prediction.  These results are helpful in resolving the long-standing puzzle of
identifying the observed P-wave state, and also shed more light on
the interpretation of the enigmatic $X(3872)$ state.  Furthermore,  this
method can also provide the explicit
 wave functions of resonances, ``compositeness" and ``elementariness"
parameters for bound states, and scattering $S$-matrix involving
these resonances \cite{Xiao:2016dsx,Xiao:2016wbs,Xiao:2016mon}, which
are rigorously obtained in the Friedrichs model and have important
applications in further studies of the resonance properties. This
scheme provides a general framework to incorporate the hadron
interaction corrections to the spectra predicted by the quark model,
and
 can be used in evaluating the other mass spectra above the
open-flavor threshold to reconcile the gaps between the quark
potential model predictions  and the experimental results.

To introduce our theoretical framework, we begin by recalling some
basic facts about the Friedrichs model. A resonance exhibiting a peak structure in the invariant mass spectrum of
the final states could be understood as a Gamow state in the famous Friedrichs model in mathematical
physics~\cite{Friedrichs:1948}. In the simplest version of the
Friedrichs model, the full Hamiltonian $H$ is
separated into the free part and the interaction part as
\bqa
H=H_0+V,
 \eqa
and the free Hamiltonian
\begin{align}
H_0=\omega_0|0\rangle\langle
0|+\int_{\omega_{th}}^\infty \omega|\omega\rangle\langle\omega|d\omega
\end{align}
has a discrete eigenstate $|0\rangle$
with eigenvalue $\omega_0>\omega_{th}$, and  continuum eigenstates
$|\omega\rangle$ with eigenvalues $\omega\in[\omega_{th},\infty)$,
$\omega_{th}$ being the threshold for the continuum states,
and they are normalized as
\bqa\label{normal}
\langle 0|0\rangle=1, \langle \omega|\omega'\rangle=\delta(\omega-\omega'),\langle 0|\omega\rangle=\langle \omega|0\rangle=0.
\eqa
The interaction part serves to couple the discrete state and the continuous state as
\bqa
V=\lambda\int_{\omega_{th}}^\infty \big[f(\omega)|\omega\rangle\langle 0
|+f^*(\omega)|0\rangle\langle \omega |\big]\mathrm{d}\omega,
\eqa
where $f(\omega)$ function denotes the coupling form factor between the discrete
state and the continuum state and $\lambda$ denotes the coupling
strength.  This eigenvalue problem for the Hamiltonian can be exactly
solved. In the Rigged-Hilbert-Space~(RHS) formulation of the quantum mechanics developed
by A.Bohm and M.Gadella, the discrete state becomes generalized
eigenstate with a
complex eigenvalue, which corresponds to the resonance state called
Gamow state~\cite{Bohm:1989,Civitarese200441}.
The
relation of the Gamow state and the pole in the scattering amplitude
in the $S$-matrix theory is also
straightforward~\cite{Civitarese200441}. By summing the
perturbation series,
 I.Prigogine and his collaborators also obtained
a similar mathematical structure~\cite{Prigogine:1991}.  Properties of
Gamow states could be represented by the zero point of $\eta(x)$
function on the unphysical sheet of the complex energy plane, where
\begin{align}
\eta^{\pm}(x)=x-\omega_0-\lambda^2\int_{\omega_{th}}^\infty\frac{|f(\omega)|^2}{x-\omega\pm
i \epsilon}\mathrm{d}\omega\,.\label{eq:eta-pm}
\end{align}
In general, when $\lambda$ increases from 0, the zero point moves away
from the real axis to the second Riemann sheet.
The wave function of the Gamow state is expressed as
\begin{eqnarray}
|z_R\rangle=N_R\Big(|0\rangle+\lambda\int_{\omega_{th}}^\infty\mathrm{d}\omega\frac{f(\omega)}{[z_R-\omega]_+}|\omega\rangle\Big),
\label{eq:Gamow-state-right}
\end{eqnarray}
and the conjuate for a pair of resonance poles on the second Rienmann sheet, where the
$[\cdots]_\pm$ means the analytical continuations of the
integration~\cite{Xiao:2016dsx}. There could also be bound-state and
virtual-state solutions both for $\omega_0>\omega_{th}$ and $\omega_0<\omega_{th}$,
with the wave function being
\begin{align}
|z_{B,V}\rangle=N_{B,V}
\Big(|0\rangle+\lambda\int_{\omega_{th}}^\infty\frac{f(\omega)}{z_{B,V}-\omega}|\omega\rangle\mathrm{d}\omega\Big),
\label{eq:Bound}
\end{align}
for a bound~(virtual) state $|z_B\rangle$~($|z_V\rangle$) at $z_B$
($z_V$) on the first~(second) Riemann sheet below the threshold, where
$N_{R,B,V}^{(*)}$ are the normalizations. For virtual states, the
integral should be continued to $z_V$ on the second sheet. A generalization
of the Friedrichs model to include multiple discrete states and
multiple continuum states is also worked out and readers are
referred to Refs.\cite{Xiao:2016dsx,Xiao:2016wbs,Xiao:2016mon} for
more detailed discussions.

Inspired from QCD one-gluon exchange interaction and the confinement,
the Godfrey-Isgur model~\cite{Godfrey:1985xj}, with partially
relativized linear confinement, Coulomb-type, and color-hyperfine
interactions, provides very successful predictions to the mass
spectra of the conventional meson states composed of $u$, $d$, $s$,
$c$, and $b$ quarks,  but its predictions with regard to the states above the
open-flavor thresholds are not as good as those below. These
discrepancies might arise from the neglecting of the coupling between these ``bare"
meson states and their decay channels~(both open and closed) as they
mentioned~\cite{Godfrey:1985xj}. In our scheme, the GI's Hamiltonian
which provides the discrete bare hadron eigenstates
can be effectively viewed as the free Hamiltonian in the Friedrichs model, and
the interactions between the bare states of $H_0$ and the continuum
states is modeled by the QPC
model~\cite{Blundell:1995ev}, which modify the spectrum above the
open-flavor thresholds.  The stronger the coupling is, the larger  the
influence is.  The wavefunctions in the QPC model is chosen to be the
same as the eigenstate solutions in the GI model which is approximated by a
combination of a set of harmonic oscillator basis.
Since the OZI-allowed channel will be more strongly coupled to the bare
states than the OZI-suppressed channel, the pole shift is dominantly caused by
the these channels. So, we include only the OZI allowed channels
in our analysis.

In the spirit of the Friedrichs model, suppose a discrete state $|0;JM\rangle$ with spin $J$,
 coupled to a continuum composed of two hadrons $|
p;JM,LS\rangle$ with a total angular momentum quantum numbers $J, M$, orbital angular
momentum quantum number $L$, total  spin $S$, the center of mass
(c.m.) momentum $p$ for
the two particles, and the reduced mass $\mu$. In the non-relativistic theory, the free Hamiltonian in the
c.m. frame  can be expressed as
\begin{align}
H_0=& M_0 \sum_{M}|0;JM\rangle\langle0; JM| \nonumber\\
+&\sum_{L, S}
\int p^2\mathrm d p\, \omega |p;JM;LS\rangle \langle p;JM;LS|
 \,.
\end{align}

 The interaction between the discrete states and the continuum states
is rotationally invariant and we can confine ourselves to a fixed $JM$
channel and omit the $JM$ indices.
The matrix elements of the interaction potentials can be expressed as
\cite{Xiao:2016mon}
\begin{align}
H_{01}=\sum_{S,L}\int d\omega  f_{SL}(\omega)|0\rangle\langle
\omega,LS|+h.c.
\end{align}
by absorbing a phase space factor $\sqrt{\mu p}$ in both  $f_{SL}(\omega)$ and
$|\omega,LS\rangle$.

The definition of the meson state is different from the one in
Ref.~\cite{Hayne:1981zy} by omitting the factor $\sqrt{2E}$ to ensure the correct normalizations.
Then, the meson coupling $A\rightarrow BC$ can be defined as the
transition matrix element
\bqa\langle BC|T|A\rangle=\delta^3(\vec{P_f}-\vec{P_i})M^{ABC}\eqa
where
the transition operator $T$ is the one in the QPC model
\bqa
T=-3\gamma\sum_m\langle 1 m 1 -m|00\rangle\int d^3\vec{p_3}d^3\vec{p_4}\delta^3(\vec{p_3}+\vec{p_4})
\nonumber\\
\times\mathcal{Y}_1^m(\frac{\vec{p_3}-\vec{p_4}}{2})\chi_{1 -m}^{34}\phi_0^{34}\omega_0^{34}b_3^\dagger(\vec{p_3})d_4^\dagger(\vec{p_4}).
\eqa

By the standard derivation one can obtain  the amplitude
$M^{ABC}$ and the partial wave amplitude $M^{SL}(P(\omega))$
as in Ref.~\cite{Blundell:1995ev}. Then the form factor $f_{SL}$ which
describes the interaction between $|A\rangle $ and $|BC\rangle$  in the
Friedrichs model can be obtained  as
\bqa
f_{SL}(\omega)=\sqrt{\mu P(\omega)}M^{SL}(P(\omega)),
\eqa
where $P(\omega)=$$\sqrt{\frac{2M_BM_C(\omega-M_B-M_C)}{M_B+M_C}}$ is
the c.m. momentum,  $M_B$ and $M_C$ being the masses of meson $B$ and
$C$ respectively.
Now, after including more continuum states, $\eta(z)$ function can be expressed as
\bqa
\eta^{\pm}(z)=z-\omega_0&-&\sum_n\int_{\omega_{th,n}}^\infty\frac{\sum_{S,L}|f_{SL}^n(\omega)|^2}{z-\omega\pm
i \epsilon}\mathrm{d}\omega,
\eqa
where $\omega_{th,n}$ denotes the energy of the $n$-th threshold.
Notice that  this function can be continued to an analytic function
$\eta(z)$ defined on a $2^n$-sheet Riemann
surface in the case of $n$ thresholds. The poles of a scattering
amplitude are just
the zeros of the $\eta(z)$ function~\cite{Xiao:2016mon}, and its
real part and imaginary part represent the mass and half-width of the
Gamow state. Only the Gamow states close to the physical region could
significantly influence the observables such as cross section or invariant
mass spectrum of the final states.

With the parameters in the GI model ~\cite{Godfrey:1985xj},
we first reproduced the results of GI by approximating the wave function of the P-wave
charmonium states and the charmed mesons with 30 Harmonic Oscillator
wave function basis. Using these wave functions of the meson
states in the QPC model, one could then obtain the
coupling form factor in the Friedrichs model. The only parameter of
the QPC model is  $\gamma$, which represents the quark pair production
strength from the vacuum. In the literature, various values of
$\gamma$
are chosen
in different situations, and a typical value is chosen as $6.9$ \cite{Barnes:2005pb,Kokoski:1985is}. However, since the wave functions used here are
different from theirs, it is no need to choose the same value as
there.   We choose it to be a value around
$4.0$ such that all the observed $P$-wave first excited charmonium state
spectrum can be reproduced well simultaneously in our
scheme.

The coupled channels are chosen up to  $D^*\bar D^*$  in the four cases.
The $\chi_{c2}(2P)$ state can couple to $D\bar D$, $D\bar D^*$, and $D^*\bar D^*$ in
both $S$ and $D$-wave. For the $\chi_{c1}(2P)$
and $h_c(2P)$ states, the coupled channels are $D\bar D^*$, and $D^*\bar D^*$.
In the case of the $\chi_{c0}(2P)$ state,  the coupled channels are $D\bar D$, and
$D^*\bar D^*$.

The poles of scattering amplitude~(zero of the $\eta(z)$ function)
could be extracted by analytically continuing $\eta(z)$  to the
closest Riemann sheet. To make this scheme more friendly to the
experimentalists, one may approximate $\eta(z)$ by a Breit-Wigner
parametrization as $\eta(z)\approx z-M_{BW}+i \Gamma_{BW}/2$, and the
mass parameter is  determined by solving
\bqa\label{massBW}
M_{BW}-\omega_0-\sum_n\mathcal{P}\int_{\omega_{th,n}}^\infty\frac{\sum_{SL}|f^n_{SL}(\omega)|^2}{M_{BW}-\omega
}\mathrm{d}\omega=0,
\eqa
on the real axis where $\mathcal{P}\int$ means principal value integration
and the Breit-Wigner partial width of the $n$-th open channel is expressed
as
\bqa\label{widBW}
\Gamma_{BW}^n=2\pi \sum_{S,L} |f_{SL}^n(M_{BW})|^2,
\eqa
and the total width $\Gamma_{tot}=\sum_{n}\Gamma_{BW}^n$.  It is
worth mentioning that this approximation, Eq.~(\ref{massBW}) and
(\ref{widBW}) together, is only valid when it is used to represent a
narrow resonance far away from the thresholds.

\begin{table}[htp]
\begin{center}
\caption{\label{compilation}Comparison of the experimental masses and
the total widths (in MeV) \cite{Olive:2016xmw} with our results.}

\begin{tabular}{|c|c|c|c|c|c|c|}
\hline
\hline
$n^{2s+1}L_J$& $M_{expt}$&$\Gamma_{expt}$ & $M_{BW}$&$\Gamma_{BW}$&pole& GI  \\
\hline
$2^3P_2$ & $3927.2\pm 2.6$ &$24\pm 6$&3920&10 &3920-4i   &$3979$    \\
\hline
$2^3P_1$ & $3942\pm 9$ &$37^{+27}_{-17}$& & & 3934-40i    &$3953$    \\
$$ & $3871.69\pm 0.17$ &$<1.2$& 3871  & 0 &3871-0i &   \\
\hline
$2^3P_0$ & $3862^{+66}_{-45}$ &$201^{+242}_{-149}$&3878&11 &3878-5i    &$3917$  \\
\hline
$2^1P_1$ & $ $ &$ $&3895& 37 & 3902-27i   &  $3956$   \\
\hline
\hline
\end{tabular}
\end{center}
\end{table}%

The numerical results of the extracted pole position and related
Breit-Wigner parameters are shown in Table~\ref{compilation}.   If
there is only one open channel, usually one Gamow state which
originates from the bare state is expected,
but sometimes there could also be an extra virtual state or bound state
generated by the form factor $f(\omega)f(\omega)^*$  when
the coupling is strong, which exhibits the molecular nature of this
state. Here, the $X(3872)$ is just of this nature.
 In Ref.~\cite{Xiao:2016dsx}, we discussed the general condition
for
this kind of virtual or bound state poles.

For the $2{}^3P_2$ channel,
$D\bar{D}$ and $D\bar{D}^*$ thresholds are open for the
$\chi_{c2}(2P)$. This pole is shifted
from the GI's value
down to about 7 MeV below the observed value. Its width is about
10 MeV, a little smaller than the observed one. The branching ratio
between $D\bar{D}$ and $D\bar{D}^*$ is 2.4, which demonstrates that
$D\bar{D}$ is its dominant decay channel. Its decay probability to
$D\bar{D}^*$ is relatively small, but there is still some possibility that there is the contribution of $X(3930)$ in the $D\bar{D}^*$ mass distribution in experiments.

In the $2^{3}P_1$ channel,  one pole is shifted down from the
bare state  to about 3934 MeV with fairly large width, while another
bound state pole emerges just below the threshold around $3871$ MeV, which is consistent with the
$X(3872)$ found in the experiment. If the coupling
strength $\gamma$ is tuned smaller, this bound-state pole will move
across the $D\bar{D}^*$ threshold to the second sheet and becomes a virtual state pole.
This pole is dynamically generated from the form factor
which is an evidence of the molecular origin of the state.
It is natural to assign this bound state pole to the $X(3872)$, and
the higher state generated from GI's bare state might be related to
the $X(3940)$ state.

\begin{figure}[h]%
\begin{center}%
\includegraphics[height=35mm]{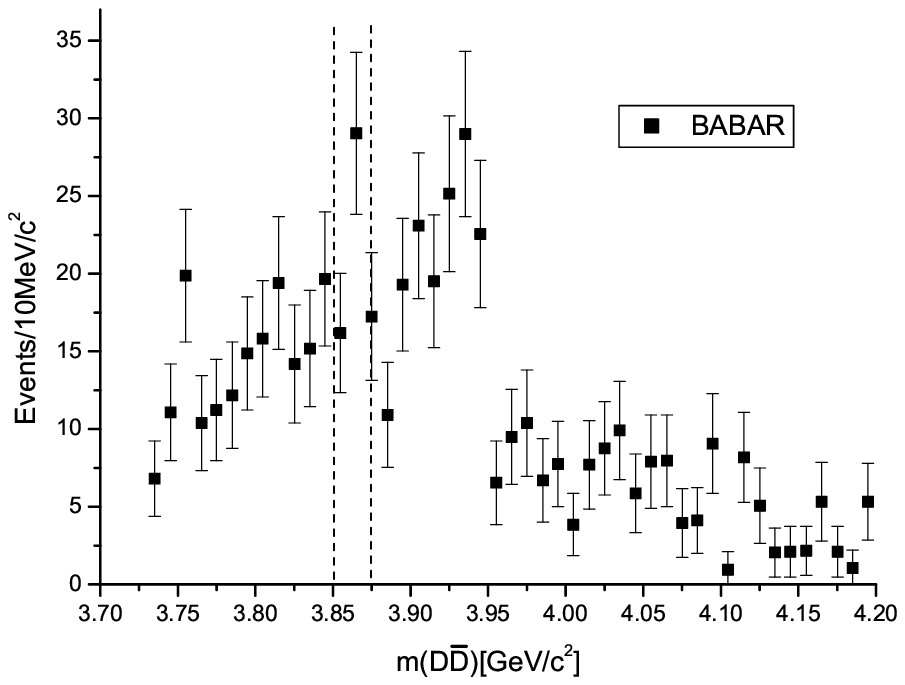}
\hspace{-0.4cm}\includegraphics[height=35mm]{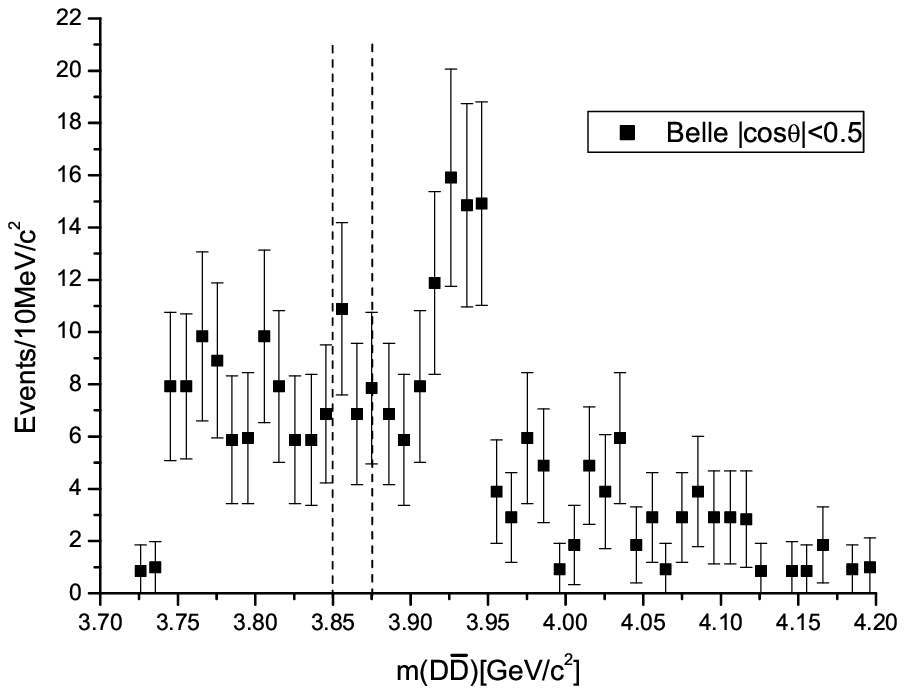}
\caption{\label{exp3860} The mass distribution of
$\gamma\gamma\rightarrow D\bar{D}$ from BABAR~\cite{Aubert:2010ab} and
Belle~\cite{Uehara:2005qd}. The data of Belle is the one for
$|\cos\theta^*|<0.5$.  The two dashed lines are set at $m(D\bar D) =3850$ MeV and 3875 MeV.}
\end{center}%
\end{figure}

In the $2{ }^3P_0$ channel, the $\chi_{c0}(2P)$ state is found to be a narrow resonance  at about
3878 MeV. We noticed that the mass of the newly observed $\chi_{c0}(2P)$
candidate is at 3862 MeV with a width $201^{+242}_{-149}$
MeV~\cite{Chilikin:2017evr}, having a
large uncertainty. Although the $D\bar{D}$ channel is
OZI-allowed for the $\chi_{c0}(2P)$ state, in this calculation we find that
the coupling between  the $D\bar{D}$ channel and  the bare
$\chi_{c0}(2P)$  is unexpectedly weak which causes the narrow width.
This narrow width is roughly only the
bin size  of the data in \cite{Aubert:2010ab, Uehara:2005qd} and
is also smaller than the one in the more recent \cite{Chilikin:2017evr}.
So, in the future experiments, we propose a further exploration with a higher
resolution in this energy region to see whether there is a narrow
signal missing in the present data. An interesting observation is that there
seems to be a simultaneous small excess at the vicinity of about 3860 MeV
in $\gamma\gamma\rightarrow D\bar{D}$ experiment of both
Belle~\cite{Uehara:2005qd} and BABAR
Collaboration~\cite{Aubert:2010ab}. Especially in BABAR's data, the
small structure extends to a dip around 3880MeV. Notice the data points in the region of $3850\mathrm{MeV}<m(D\bar D)<3875\mathrm{MeV}$ in Fig.~\ref{exp3860}.

The pole mass of $h_c(2P)$ state is predicted at around
3902 MeV in this
scheme. As we have mentioned, $\chi_{c2}(2P)$, $\chi_{c1}(2P)$,
$h_c(2P)$  all
couple to the $D \bar D^*$ channel, which has no definite
$C$-parity. This means that the enhancement above the $D\bar D^*$ threshold
contains all the contributions from these states. To detect the
$h_c(2P)$ signal, one needs to look for it in a negative $C$-parity channel such
as $\eta_c \gamma$ in this energy region.

Further remarks about the $X(3872)$ is in order. In our calculation,
although it is found to be a
bound state, we can not exclude the possibility of  a virtual
state nature~\cite{Hanhart:2007yq,Kang:2016jxw}, since only a small shift down of the $\gamma$ parameter
will move it to the second sheet. In \cite{Zhou:2013ada}, improving
the approach adopted in \cite{Pennington:2007xr},  a
dispersion relation method combined with the QPC model is also used in discussing
the charmonium-like states, where $X(3872)$ can also be produced. However,
since the wave function for the $X(3872)$ can not be obtained there and
the wave function used in the QPC model there is inaccurate, further
discussion on the nature of the $X(3872)$ may not be accurate. In
the present scheme, the exactly solvable Friedrichs model provides a more solid
theoretical setup and since the more accurate hadron wave function of the GI model is
used in the QPC model, the result here would more accurately
discribe the nature of the $X(3872)$. Moreover, since the wave function for  the
$X(3872)$ can be rigoroursly solved in terms of the discrete state and
the continuum states, one can also find out its compositeness and
elementariness. The compositeness of a bound state, defined as the
probability of finding the $n$-th continuous states in the bound
state, is expressed in the Friedrichs model as
\bqa
X_n=\frac{1}{N^2}\int_{\omega_{th,n}}^\infty\mathrm{d}\omega\sum_{S,L}\frac{|f_{SL}^n(\omega)|^2}{(m_B-\omega)^2},
\eqa
where the normalization factor is
\bqa
N=(1+\sum_n\int_{\omega_{th,n}}^\infty\mathrm{d}\omega\sum_{S,L}\frac{|f_{SL}^n(\omega)|^2}{(m_B-\omega)^2})^{1/2}.
\eqa
If the $X(3872)$ is a bound state, the relative ratio of finding
$c\bar{c}$ and $D\bar{D}^*$ in the state is about $1:3.1\sim 1:9.3$ if we
tune the $\gamma$ parameter such that the $X(3872)$ locates within
$3.8710\sim 3.8717$, showing the
dominance of the continuum part in this state,  which also demonstrates
its molecular dominant
nature~\cite{Tornqvist:1993ng,Gamermann:2009uq,Guo:2013nza,Meng:2013gga}.
Since the position of the $X(3872)$ is very close to the threshold,
the compositeness is sensitive to its position.
 In comparison, in \cite{Meng:2013gga}, by analyzing the
production rate of CMS~\cite{Chatrchyan:2013cld} and
CDF~\cite{Acosta:2003zx} data within the framework of NRQCD
factorization, the  $c\bar c$ component is estimated to be
$22-44\%$, which is consistent with our value.   However, our result is different from
\cite{Takizawa:2012hy}, in which the $c\bar c$ component is about $6\%$. Nevertheless, both
results favor a large $D\bar D^*$ component. Another QCD sum-rule
analysis~\cite{Matheus:2009vq} predicts a larger $c\bar c$ component, about $97\%$, but the mass of
the state is too low, at around 3.77 GeV, compared to the observed one of $X(3872)$.
It is
worth emphasizing that for a resonance, the
compositeness and elementariness parameters will become complex
numbers~\cite{Xiao:2016mon}, so they have no rigourous
definitions, but some definitions proposed in the literature~\cite{Sekihara:2014kya,Guo:2015daa} might be able to approximately
describe these quantities.

In this paper, using the exactly solvable Friedrichs model, we
propose a general framework
to include the hadron interaction corrections to the quark model
spectrum predictions, in particular to the generally accepted GI's standard
results. The explicit wave function for the resonances can be obtained
and the compositeness and elementariness of the bound states can be
calculated which are important for further study of the properties
of the state. Using this scheme, we could reproduce the first excited
P-wave charmonium-like states. In particular, we find that the $X(3872)$
could be dynamically generated in a natural way by the
coupling of the bare $\chi_{c1}(2P)$ state and continuum states, but
its continuum 
components is larger. The $\chi_{c0}(2P)$ is found  unexpectedly to be a narrow one. We also predict the appearance of the $h_c(2P)$
state to be at about 3902 MeV with a pole width of about 54MeV. This scheme is
promising in matching the predictions of GI model with the
observed states. The acceptable consistency of our results and
experiments means that the hadron interactions really give large corrections to
the GI's results for open flavor channels which can reconcile the
descrepancy between the quark model prediction and the experiments.

\begin{acknowledgments}
Helpful discussions with Dian-Yong Chen, Hai-Qing Zhou, Ce Meng, and Xiao-Hai Liu are appreciated.
Z.X. is supported by China National Natural
Science Foundation under contract No.  11105138, 11575177 and 11235010. Z.Z is supported by the Natural Science Foundation of Jiangsu Province under contract No. BK20171349.
\end{acknowledgments}

\bibliographystyle{apsrev4-1}
\bibliography{Ref}

\end{document}